
\magnification=1200 \vsize=25truecm \hsize=16truecm \baselineskip=0.6truecm
\parindent=1truecm \nopagenumbers \font\scap=cmcsc10 \hfuzz=0.8truecm
\def\xup{\overline x}
\def\xdo{\underline x}
\def\zup{\overline z}
\def\zdo{\underline z}
\def\yup{\overline y}
\def\ydo{\underline y}
\def\Aup{\overline A}
\def\Ado{\underline A}

\null \bigskip  \centerline{\bf LINEARIZATION AND SOLUTIONS}
\centerline{\bf OF THE DISCRETE PAINLEV\'E-III EQUATION}
\vskip 2truecm
\centerline{\scap B. Grammaticos}
\centerline{\sl LPN, Universit\'e Paris VII}
\centerline{\sl Tour 24-14, 5${}^{\grave eme}$\'etage}
\centerline{\sl 75251 Paris, France}
\bigskip
\centerline{{\scap F.W. Nijhoff}}
\centerline{\sl Department of Mathematics and Computer Science}
\centerline{\sl Clarkson University, Potsdam NY 13699-5815, USA}
\bigskip
\centerline{{\scap V. Papageorgiou}}
\centerline{\sl LPN, Universit\'e Paris VII}
\centerline{\sl Tour 24-14, 5${}^{\grave eme}$\'etage, 75251 Paris, France}
\centerline{\sl and Department of Mathematics and Computer Science}
\centerline{\sl Clarkson University, Potsdam NY 13699-5815, USA}
\bigskip
\centerline{\scap A. Ramani}
\centerline{\sl CPT, Ecole Polytechnique}
\centerline{\sl CNRS, UPR 14}
\centerline{\sl 91128 Palaiseau, France}
\bigskip
\centerline{{\scap J. Satsuma}}
\centerline{\sl Department of Mathematical Sciences}
\centerline{\sl University of Tokyo}
\centerline{\sl 3-8-1 Komaba, Meguro-ku, Tokyo 153, Japan}

\vskip 3truecm \noindent
Abstract \smallskip
\noindent
We present particular solutions of the discrete Painlev\'e III
(d-P$\rm_{III}$) equation of rational and special function (Bessel)
type. These solutions allow us to establish a close parallel between
this discrete equation and its continuous counterpart. Moreover, we
propose an alternate form for d-P$\rm_{III}$ and confirm its
integrability by explicitly deriving its Lax pair.

\vfill\eject

\footline={\hfill\folio} \pageno=2

{\scap 1. Introduction}
\smallskip
\noindent

The discrete Painlev\'e III (d-P$\rm_{III}$) equation holds a
particular position among all the discrete Painlev\'e equations. It
was the first transcendent that was not discovered ``accidentally'' in
some physical application [1], but, instead, was derived ``on
request'' [2] using the method of singularity confinement [3]. Given
the method of derivation, the Lax pair of d-P$\rm_{III}$ was not
initially known. It was obtained shortly afterwards [4], confirming
thus the integrability of this equation and consolidating the
singularity confinement approach as integrability detector. Another
interesting feature of d-P$\rm_{III}$ is that its Lax pair is {\sl
not} of the ``usual'' differential-difference kind but, rather a
$q$-difference one of the form: $\Phi_n(qh)=L_n(h)\Phi_n(h)$,
$\Phi_{n+1}(h)=M_n(h)\Phi_n(h)$. This is not a mere novelty but has
important implications. Thus when quantizing this discrete Painlev\'e
equations [5] we discovered that the consistent quantization rule was
of Weyl, rather than Heisenberg, type i.e. $x\xup =q\xup x$ where
$x\equiv x_n$, $\xup\equiv x_{n+1}$. Apart from this ``natural''
appearance of the quantum-line relations, the quantization of
d-P$\rm_{III}$ revealed interesting factorization properties that will
be used in what follows. We are going, in fact, to produce special,
Bessel-type, solutions to d-P$\rm_{III}$ starting from these
factorizable forms.  The continuous P$\rm_{III}$ equation is also
special in the sense that it is traditionnaly given under two
different forms [6]. Although one can go from one to the other through
some independent variable transformation both are considered, in a
sense, canonical. In the discrete case, only one of these two forms
was known up to now. As we will show in this paper, the second
canonical form exists as well and moreover its Lax pair can be
obtained by generalizing the results of Joshi and collaborators [7] on
the discrete d-P$\rm_{II}$ equation.

 \medskip
{\scap 2. The first form of the discrete P$\rm_{III}$}
\smallskip  \noindent
Let us first review briefly what is known about d-P$\rm_{III}$. In
[2], we have obtained its form as:
$$\xup \xdo={\nu x^2-b\mu^n x-d\mu^{2n} \over c x^2+ax+\nu} \eqno(1)$$
The continuous limit is best obtained if we start with a change of variable
$y_n=\mu^{-n/2}x_n$. Equation (1) is transformed to:
$$\yup \ydo={\nu y^2-b\mu^{n/2}y- d \mu^n \over c \mu^n y^2+a\mu^{n/2}y+\nu}
\eqno(2)$$
and the continuous limit is obtained by taking to $\nu=-1/\epsilon^2$ and
$\epsilon
\to 0$. One must take simultaneously $\mu=1+2\epsilon$ in which case
$\mu^{n/2}$ goes over to $e^z$
leading to: $$x''={x'^2\over x}+e^z(ax^2+b)+e^{2z}(cx^3+{d\over x}) \eqno(3)$$
The Lax pair of (1) was given in [4].  We implement
the compatibility condition as: $$M_n(qh)L_n(h)=L_{n+1}(h)M_n(h)\eqno(4)$$
with $q=\alpha^2$ and
$$L=\pmatrix{\lambda_1&\lambda_1+{\kappa \over y}&{\kappa \over y}&0\cr
 0&\lambda_2&\lambda_2+\underline y&\underline y\cr
hy&0&\lambda_3&\lambda_3+y\cr
h(\lambda_4+\alpha {\kappa \over y})&h\alpha {\kappa \over y}&0&\lambda_4\cr}
\eqno(5)$$
$$M=\pmatrix{{(\alpha\lambda_1-\lambda_4)y\over \lambda_4y+\alpha \kappa}&
{\lambda_1y+\kappa \over \lambda_2y+\kappa }&0&0\cr 0&0&1&0\cr
0&0&{\lambda_3-q\lambda_2\over y+q \lambda_2}&{y+\lambda_3\over y+\lambda_4}\cr
h&0&0&0\cr}$$
where $\lambda_1=cnst.$, $\lambda_3=cnst.$,
$\lambda_2=\lambda\alpha^{n-1}$, $\lambda_4=\lambda\alpha^{n}$, $\kappa =C
\alpha^n$.

We obtain d-P$\rm_{III}$ in the form [4,5]:
$$\yup \ydo={\alpha\kappa (y+\lambda_3)(\kappa +\lambda_2 y)
\over (\kappa +\lambda_1 y) (y+\lambda_4)} \eqno(6)$$

Equation (6) is just (2) with $\mu =1/\alpha$ and the remaining
parameters easily identifiable. The important point is that (6) is
written in a factorized form. This suggests that the right-hand-side
of (6) can be considered as a product of two homographic mappings, and
this is precisely the key for the obtention of special solutions. It
is based on the fact that the homographic mapping is the discrete form
of the Riccati equation [8] and that the special solutions of the
continuous Painlev\'e equations are obtained through linearizations
via Riccati equations [9].

The simplest way to find these solutions is to assume a form for
the homographic mapping:
$$\xup =-{\alpha x+\beta\over\gamma x+\delta}\eqno(7a)$$
and thus
$$\xdo =-{\underline\delta x+\underline\beta \over \underline\gamma
x+\underline\alpha}\eqno(7b)$$
 and substitute in (1) (rather than (6)). Moreover we can freely choose
the normalization for
the mapping (7) by taking $\gamma =1$. Expressions become much simpler
if instead of working
with (1) we take $c=1$ (by simple division):
$$\xup \xdo={\nu x^2-b\mu^n x-d\mu^{2n} \over x^2+ax+\nu} \eqno(8)$$
It is well known that one can freely scale-out two of the parameters of
P$\rm_{III}$ and the same
is true for d-P$\rm_{III}$. Equating terms between (8) and the product of
(7a)(7b) we find:
$$\delta+\underline \alpha =a$$
$$\delta\underline \alpha =\alpha\underline \delta =\nu\eqno(9)$$
$$\beta\underline \delta+\underline\beta\alpha=-b\mu^n$$
$$\beta\underline \beta=-d\mu^{2n}$$
Hence, $\alpha$ and $\delta$ are constant with sum equal to $a$ and
product $\nu$ and $\beta
=\beta_0\mu^n$, with $\beta_0=\sqrt{-d\mu}$ (which can be scaled to 1).
Moreover, one constraint must be satisfied:
$$\delta+{\alpha\over\mu}=b\eqno(10)$$
Substituting back into (7a) we find:
$$\xup =-{\alpha x+\mu^n\over x+\delta}\eqno(11)$$
that can be linearized by the substitution $x=P/Q$. We find thus:
$$\overline P=-\alpha P-\mu^n Q$$
$$\overline Q=P+\delta Q\eqno(12)$$
and finally:
$$\overline{\overline Q}+(\alpha-\delta)\overline Q+(\mu^n
-\alpha\delta)Q=0\eqno(13)$$
This is just a discrete form of Bessel's equation. In fact, it is
essentially the same as the $q$-Bessel function that was derived in
[10] in relation to the $q$-discrete Toda model and which can be also
found in the classical monograph of Exton [11].  The continuous limit
of (13) is just Bessel's equation. First, we introduce
$Q=\epsilon^{-n}R$ and rewrite (13) as:
$$\overline{\overline R}+\epsilon (\alpha-\delta)\overline R+ \epsilon^2(\mu^n
-\alpha\delta)R=0\eqno(14)$$

Next, we expand $\overline R,\overline{\overline R}$ up to second
order in $\epsilon$ and taking into account that $\alpha\delta\approx-
{1\over\epsilon^2}$, we obtain in the limit $\epsilon\to 0$:
$$R''+(e^{2z}-{a^2\over 4})R=0\eqno(15)$$ i.e. precisely Bessel's
equation for $J_{a\over 2}(e^z)$ [12]. From this basic solution one
can construct higher ones, just as in the case of d-P$\rm_{II}$ [13]
in close parallel to the continuous P$\rm_{III}$. Bessel function
solutions are not the only elementary solutions for d-P$\rm_{III}$.
Rational solutions exist as well. For example, when $d=0$ we find
readily $x=0$. Other rational solutions can be found also: when $d=0,
b=0$ we obtain $x=-a/c$. As in the continuous case, a duality under
$x\to 1/x$ exists, resulting in an exchange of $(b,d)$ and $(a,c)$.
Thus, when $a=c=0$ we find $x=-{d\over b}\mu^n$ as a solution.  This a
very particular case of the general solution that can be obtained in
closed form for $a=c=0$.  Putting $x=X\mu^n$ we can reduce (1) to the
autonomous form: $\overline X\underline X=X^2-bX-d$ (where we have put
$\nu=1$ by simple division). The solution is found by a
straightforward calculation: $X=A\lambda^n+B\lambda^{-n}+C$ where
$C(\sqrt \lambda-1/\sqrt\lambda )^2+b=0$ and $AB(\lambda-1/\lambda
)^2+bC+d=0$. Thus $\lambda$ and one of the $A,B$ are free in general.
In the particular case $b=0$ we have $C=0$ but the solution involves
still two free constants.

Thus d-P$\rm_{III}$ exhibits all the richness, as far as its particular
solutions are concerned, as
its continuous counterpart.

\medskip
{\scap 3. An alternate form of the discrete P$\rm_{III}$}
\smallskip
\noindent As we explained in the introduction, the continuous P$\rm_{III}$
is given usually under
two different canonical forms and even, sometimes, a third one. So, besides
P$\rm_{III}$ written
in the form (3) we have also:
$$x''={x'^2\over x}-{x'\over z}+{1\over z}(ax^2+b)+c x^3+{d\over x}\eqno(16)$$
and
$$x''={x'^2\over x}-{x'\over z}+{1\over 4z^2}(ax^2+c x^3)+{b\over 4z}+
{d\over 4x}\eqno(17)$$
All these forms (3,16,17) are, of course, equivalent within a simple
change of variables. For
example, (17) and (16) are related through the transformation $z\to z^2$
and $x\to zx$. As we
have pointed out in section 2, the discrete form of P$\rm_{III}$ corresponds
naturally to the
form (3). This does not mean that (3) is the only continuous limit of (2).
Indeed one can
recover any form of P$\rm_{III}$ by taking the continuous limit in the
appropriate way.

Usually, one relates the continuous variable $z$ to the discrete one $n$
through some constant-step
discretization $z=n\epsilon$ and thus $\overline z  -z=z- \underline
z=\epsilon$. Changes in the
discrete independent variable can be represented by a  nonconstant step
in $z$. We start from:
$$\overline z  -z=\epsilon f_+$$
$$z- \underline z=\epsilon f_-\eqno(18)$$
where $f_+$ and $f_-$ are slowly varying functions (in particular,
$f_+-f_-= {\cal O}(\epsilon)$,
and thus $f_+^2+f_-^2=2f_+f_-+{\cal O}(\epsilon^2)$). We have:
$$\overline x=x+ \epsilon f_+ x'+ \epsilon^2 f_+^2 x''/2+ \dots$$
$$\underline x=x- \epsilon f_- x'+ \epsilon^2 f_-^2 x''/2+\dots \eqno(19)$$
Equation (1) becomes:
$$\eqalign{\xup\xdo -x^2&=\epsilon^2[(xx''-x'^2)f_+f_-+xx'{f_+-f_-\over
\epsilon}
+{\cal O}(\epsilon^3)]\cr
&=\epsilon^2({1\over \nu})(cx^4+ax^3+b\mu^nx+d\mu^{2n})+{\cal
O}({1\over\nu^2})\cr}\eqno(20)$$
and with $\nu=-{1\over \epsilon^2}$ we can obtain various
forms of P$\rm_{III}$ in the continuous limit. In order to find precisely
P$\rm_{III}$$'$
in the form (17) we take $z=(1+\epsilon)^n$, $f_+=z$, $f_-={z\over 1+
\epsilon}$, leading
to  ${f_+-f_-\over\epsilon}={z\over 1+\epsilon}$  and we choose $\epsilon$
so as to have
$\mu^n=z$. Had we wished to find P$\rm_{III}$, eq.(16), we should have
started from
eq.(2) for $y$ and then take $\epsilon$ so that $\mu^{n/2}=z$.

However this limiting procedure, although interesting, may appear somewhat
artificial since it
incorporates the change of variables that we need in order to find the
desired form of
P$\rm_{III}$. Thus one may wonder whether there exists a ``natural''
discrete analog to
P$\rm_{III}$ (16) or P$\rm_{III}$$'$ (17). Interestingly, the answer is:
yes. The hint lies in
the results we obtained in [14] where the mapping:
$${z+\zup\over x+\xup}+{z+\zdo\over x+\xdo}=
k(1+{1\over x^2})+{2z\over x}\eqno(21)$$
was identified as a discrete form of P$\rm_{III}$ (although with
restricted coefficients). Its
continuous limit can be obtained through $k = \epsilon^2  a$, and at the
limit $\epsilon \to 0$ we find: $$x''={x'^2\over x}-{x'\over z}-
{2a\over z}(x^2+1)\eqno(22)$$
Equation (21) can also be rewritten as a system,
whereupon one realizes that the missing coefficients, that would lead
to a full P$\rm_{III}$ can be
easily introduced. The final result is:
$$x+\xup ={\zeta y+\theta\over y^2-1}\eqno(23a)$$
$$y+\ydo ={\eta x+\kappa \over x^2-1}\eqno(23b)$$
with constant $\theta$ and $\kappa$, and $\zeta$, $\eta$ related through
$2\zeta=\eta+\overline\eta$, $2\eta=\zeta+\underline\zeta$. Thus $\zeta$
and $\eta$ are linear
in $n$ and staggered i.e. $\eta(n)=\zeta(n-1/2)$.

This is a remarkable result, since each of the equations of the system has
the form of
d-P$\rm_{II}$. In fact, this form has been known from the outset [2] since
the general form of
d-P$\rm_{II}$ has been obtained as:
$$\xdo+\xup ={(\alpha n+\beta )x+\gamma +\delta(-1)^n\over x^2-1}\eqno(24)$$
where the $\delta (-1)^n$ term indicates an even-odd dependence that can
be used in order to write
system (23a,b). However in these first treatments of discrete Painlev\'e
equations the even-odd
dependence was discarded since it was considered that it should disappear
in the continuous limit.
While this is true when one considers a single equation, this is not so for
the full system.
Indeed we will show that (23a,b) leads to P$\rm_{III}$ in the form (17).
First we introduce a
more convenient scaling:
$$x+\xup ={zy+\epsilon^2 a\over y^2-\epsilon^2 c}\eqno(25a)$$
$$y+\ydo ={(z-\epsilon/2)x+\epsilon^2 b\over x^2-\epsilon^2 d}\eqno(25b)$$
whereupon the continuous limit becomes straightforward. We take:
$$y={z\over 2x}-{zx'\over 4x^2}\epsilon + ({x'\over 8x^2}+{b\over 2x^2}+
{dz\over 2x^3})
\epsilon^2\eqno(26)$$
and find for $x$ (at the limit $\epsilon \to 0$)
$$x''={x'^2\over x}-{x'\over z}+{1\over z^2}(16c x^3+8ax^2)-{4b\over z}-
{4d\over  x}\eqno(27)$$
precisely the equation P$\rm_{III}$$'$, eq.(17) up to an unimportant
rescaling of $a$, $b$, $c$
and $d$. While this is an interesting result in itself, i.e. that
d-P$\rm_{II}$ can be extended
so as to  give d-P$\rm_{III}$, we shall go one step further and show
that the Lax pair for (25)
can also be obtained in the same spirit. In fact the Lax pair for
d-P$\rm_{II}$ has been obtained
in [4] and [15] but what will interest us particularly is the one obtained
by Joshi and
collaborators [7] in the framework of the discrete AKNS theory [16].
In [7] it was shown that
the discrete problem :
$${\partial \Phi_n\over \partial\zeta}=M_n(\zeta)\Phi_n,
\Phi_{n+1}=L_n(\zeta)\Phi_n\eqno(28)$$ where
$$L=\pmatrix{\zeta&x\cr x&{1\over \zeta}\cr}$$
$$M=\pmatrix{A&B\cr C&-A\cr}\eqno(29)$$
starting from simple ans\"atze for $A$, $B$ and $C$,  has a solution of
the form
$$A={\kappa\over\zeta^3}+{-2\kappa x\underline x+n+\nu\over\zeta}+
{\kappa\zeta}\eqno(30)$$
$$B=-{2\kappa\xdo\over\zeta^2}+2\kappa x,\quad C=-{2\kappa x\over\zeta^2}+
2\kappa \xdo$$
with $\kappa$, $\nu$ constants, leading to d-P$\rm_{II}$
$$\xdo+\xup={(\alpha n +\beta)x\over x^2-1}\eqno(31)$$ with $\alpha=
{1\over\kappa}$,
$\beta={(\nu+1/2)\over\kappa}$.

This is not, however, the {\sl most general} solution of this AKNS problem
that can be reached
with the same ansatz for $A$, namely $A={p\over\zeta^3}+{q\over\zeta}+r\zeta$,
(without
fixing the form of $B$ and $C$). Eliminating $B$ and $C$ in terms of $A$ we
obtain an equation
for the latter that reads: $$\displaylines{\quad\zeta^2\xup\xdo (A -\Aup)+
\zeta \xup\xdo
+x\xup (A(\xdo^2+1)+\Ado (\xdo^2-1)) -x\xdo (\Aup (\xup^2+1)+{\overline{\Aup}}
(\xup^2-1))\hfill\cr\hfill  -x(\xup+\xdo)\zeta^{-1}
+\xup\xdo (A-\Aup )\zeta^{-2}+\xup\xdo\zeta^{-3}=0\quad(32)\cr}$$
The solution for $A$ is exactly the same as in Eq. (30) (but $B$ and $C$ are
more complicated).
The equation for $x$ can be written as $$\overline W -\underline W=0\qquad
{\rm where}\qquad
W=(\xup+\xdo)(x^2-1)-x(\alpha n+\beta)\eqno (33)$$
Integrating (33) we find $W=\gamma +\delta (-1)^n$ leading to
exactly equation (24) above which, as we saw, is equivalent to d-P$\rm_{III}$,
Eq. (23). The
result of reference [7] corresponds to taking $W=0$ as a solution of (33),
which is the only
choice leading to simple expressions for $B$ and $C$.

Thus the alternate discrete d-P$\rm_{III}$ equation has a Lax pair, a fact
that establishes its
integrability. Contrary to (5), this eigenvalue problem is a standard one
and not of
$q$-difference type.

\smallskip
{\scap 4. Conclusion}
\smallskip
\noindent
In the previous sections we have shown that the discrete P$\rm_{III}$
equation is the perfect analog to its continuous counterpart. Not only
does it possess special solutions (in terms of rational or discrete
Bessel functions) but, also, it exists under two different canonical
forms that correspond to the two canonical forms of the continuous
P$\rm_{III}$.  Furthermore, from the case of $q$-deformed P$\rm_{III}$
(4-6), we conjecture the existence of an interesting class of new
special functions which are the $q$-deformed difference analogs of the
P$\rm_{III}$ transcendent.

The fact that most features of integrable continuous equations can be
extended to the discrete case offers us a handle on integrable
discrete systems. Following the discovery of the singularity
confinement property (that allows one to detect integrable discrete
systems) the fact that there exists a continuous/discrete parallel
opens a new road in the study of discrete integrability.

\smallskip
{\scap Acknowledgements}
\noindent The authors have benefited from stimulating discussions with
K.M. Tamizhmani.

\smallskip
{\scap References}.
\smallskip
\item{[1]}      E. Br\'ezin and V.A. Kazakov, Phys. Lett. 236B (1990) 144
V. Periwal and D. Shevitz, Phys. Rev. Lett. 64 (1990) 1326.
\item{[2]} A. Ramani, B. Grammaticos and J. Hietarinta, Phys. Rev. Lett. 67
(1991) 1829.
\item{[3]} B. Grammaticos, A. Ramani and V. Papageorgiou, Phys. Rev. Lett. 67
(1991) 1825.
\item{[4]} V.G. Papageorgiou, F.W. Nijhoff, B. Grammaticos and A. Ramani,
Phys. Lett. A164 (1992) 57.
\item{[5]}  B. Grammaticos, A. Ramani, V. Papageorgiou  and F. Nijhoff
        Jour. Phys. A 25 (1992) 6419.
\item{[6]} E.L. Ince, {\sl Ordinary differential equations},
(Dover, N.Y. 1956).
\item{[7]} N. Joshi, D. Bartonclay, R.G. Halburd, Lett. Math. Phys. 26
(1992) 123
\item{[8]} A. Ramani, B. Grammaticos and G. Karra, Physica A 81 (1992) 115.
\item{[9]} V.A. Gromak and N.A. Lukashevich, {\sl The analytic solutions of the
Painlev\'e equations}, (Universitetskoye Publishers, Minsk 1990), in Russian.
\item{[10]} K. Kajiwara and J. Satsuma, J. Phys. Soc. Japan 60 (1991) 3986.
\item{[11]} H. Exton, {\sl $q$-Hypergeometric functions and applications},
(Ellis Harwood Ltd.
1983).
 \item{[12]} M. Abramowitz and I. Stegun, {\sl Handbook of Mathematical
Functions}, (Dover N.Y. 1972).
\item{[13]} A. Ramani and B. Grammaticos, Jour. Phys. A 25 (1992) L633.
\item{[14]}     A.S. Fokas,  A. Ramani, B. Grammaticos, J. Math. An. Appl.
to appear.
\item{[15]} F.W. Nijhoff and V.G. Papageorgiou, Phys. Lett. 153A (1991) 337.
\item{[16]} M.J. Ablowitz and J. Ladik, J. Math. Phys. 17 (1976) 1011.

\end